\documentclass{kluwer}    
\usepackage{graphicx}
\newdisplay{guess}{Conjecture}

\begin{document}                                                                                   
\begin{article}
\begin{opening}

\title{Gas metallicities and early evolution of distant radio galaxies.}
\author{Villar-Mart\'\i n M.$^1$, Fosbury R.$^2$, Vernet J.$^3$, Cohen M.$^4$, Cimatti A.$^5$, di Serego Alighieri S.$^5$ } 
\runningauthor{M. Villar-Mart\'\i n}
\runningtitle{High redshift radio galaxies}
\institute{$^1$Univ. Hertfordshire (UK); $^2$ST-ECF (Germany); $^3$ESO (Germany); $^4$ Caltech (USA); $^5$Arcetri (Italy)}
\date{\today}

\begin{abstract}

	By modeling  the rich emission line spectra of a sample of high redshift
(HzRG, $z\sim$2.5) radio galaxies we find  that solar and supersolar metallicities
are common in the extended gas of these objects. Our models and the comparison with 
 high redshift quasars suggest that HzRG at $z\sim$2.5 are associated with intense star formation activity. This is consistent with chemical 
evolution models for giant ellipticals and it supports
the idea that distant powerful radio galaxies are progenitors of giant ellipticals.
We might be witnessing different evolutive
status in different objects.
\end{abstract}

\end{opening}           

\section{Introduction}  

High redshift radio galaxies ($z>$2, HzRG) are believed to be progenitors
of giant ellipticals (CDs) (Best et al. 1998, McLure \& Dunlop 2000). The study of 
the early stages of the formation and evolution of these massive
(proto-)galaxies is of primary importance  to understand galaxy
formation scenarios.

 Some of the important questions about
HzRG concern the 
evolutionary status of the underlying galaxy and the connection between the formation
of the galaxy and
the central black hole. Is the  host galaxy 
fully formed yet? Is there an underlying old stellar population or maybe the
galaxy has not even formed  the bulk of its stars yet? 

It has been discussed  in this meeting how metal abundances
(constrained from the SED, absorption and/or emission lines)
can be used as probes of star formation and galaxy evolution. 
We show in this paper how we have used the emission line spectra 
(in particular, the NV$\lambda$1240 line) 
of a sample of HzRG to constrain the gas metal abundances and 
the conclusions we draw  about the evolutionary status of these galaxies.

\section{The data and the modeling code}

The spectra of 9 HzRG (2.3$\leq z\leq$3.6) were obtained 
with the Low Resolution Imaging Spectrometer  at the 
Keck II telescope. Detailed description of the sample, observing runs and 
data 
reduction will be presented in Vernet et al. (2001, in prep.). See also Fosbury et al. (1999).

We used the multipurpose code Mappings Ic developed by Luc
Binette. See (Villar-Mart\'\i n et al., 1999) for the modeling method.

\section{The ``NV diagram''}

Hamann and Ferland (1993, 1999) (HF93, HF99)
showed that  high redshift quasars ($z>$2) define a tight correlation on
the diagnostic diagram NV$\lambda$1240/HeII$\lambda$1640 vs. NV/CIV$\lambda$1550. 
The modeling of the emission line ratios lead the authors to 
conclude that the two NV ratios imply supersolar metallicities in
the broad line region of many  high redshift quasars. They 
interpret the correlation in the NV diagram as 
a sequence in metallicity such that the highest redshift/most luminous
objects show the highest metallicities ($\geq$10$\times Z\odot$).

	When we plot  the HzRG of our sample in the NV diagram,
we were surprised to find that the radiogalaxies  define a correlation 
 parallel to the quasar line (see Fig.1, top left diagram).  Our first idea was that
we are witnessing, as for distant quasars, different levels of  metal enrichment of the gas  
 from object to object and, maybe, supersolar metallicities.
An important difference with quasars is that we are talking about the
{\it narrow line gas} (extended over several tens of kpc) rather than the 
broad lines gas (very close to
the nucleus) studied by HF.

\section{The models: Results}

In order to test the validity of this interpretation it was first necessary
 to explore whether  models other than a metallicity sequence could 
reproduce the
observations: Villar-Mart\'\i n et al. (1999) studied the effects of shock ionization
(vs. active nucleus (AGN)  photoionization), 
and the influence of the AGN continuum shape,  density 
and/or ionization
parameter (U)\footnote{U is the quotient of the density of ionizing photons incident
on the gas and the gas density: 
$U=\int_{\nu_0}^{oo}{\frac{f_{\nu} d\nu / h\nu}{c n_H}}$}. 
We showed that  these models could not explain neither the NV correlation, neither
the very strong NV emission observed in some objects.

	We then investigated whether a  metallicity sequence can
 explain the NV behaviour {\it and  be consistent with
the other emission line ratios}. Thanks to the high S/N of the spectra, we could use
 many emission
lines (never detected before in HzRG) to test our models. We assumed that the gas 
(100 cm$^{-3}$) is photoionized
by a power law of index $\alpha$=-1.0 \cite{villar99} and the same U=0.035 
for all the objects (suggested by the little variation of 
CIV/CIII] and CIV/HeII, Fig.1).

 We found that (see Fig.1, top left diagram):

* A sequence in metallicity 
can reproduce  both the observed correlation and the strength of the NV emission.
The heavy element abundances relative to H vary between 0.4 and 4 $\times Z\odot$ 

*  The N abundance increases quadratically instead of linearly
	
*  There is good agreement between the model  predictions  and 
the data  {\it in most diagrams}. NIV]$\lambda$1488 is a problem. It is predicted
to be stronger than observed. Non of the models we explored  can
explain this discrepancy. A similar inconsistency has been
 reported  for the Seyfert galaxy NGC1068 \cite{kra00}. However,
the fact that both the data  and the models define a tight correlation in
the NIV diagram (see also OIII] diagram) supports a metallicity sequence.

\begin{figure} 
\centerline{\includegraphics[width=4.7in]{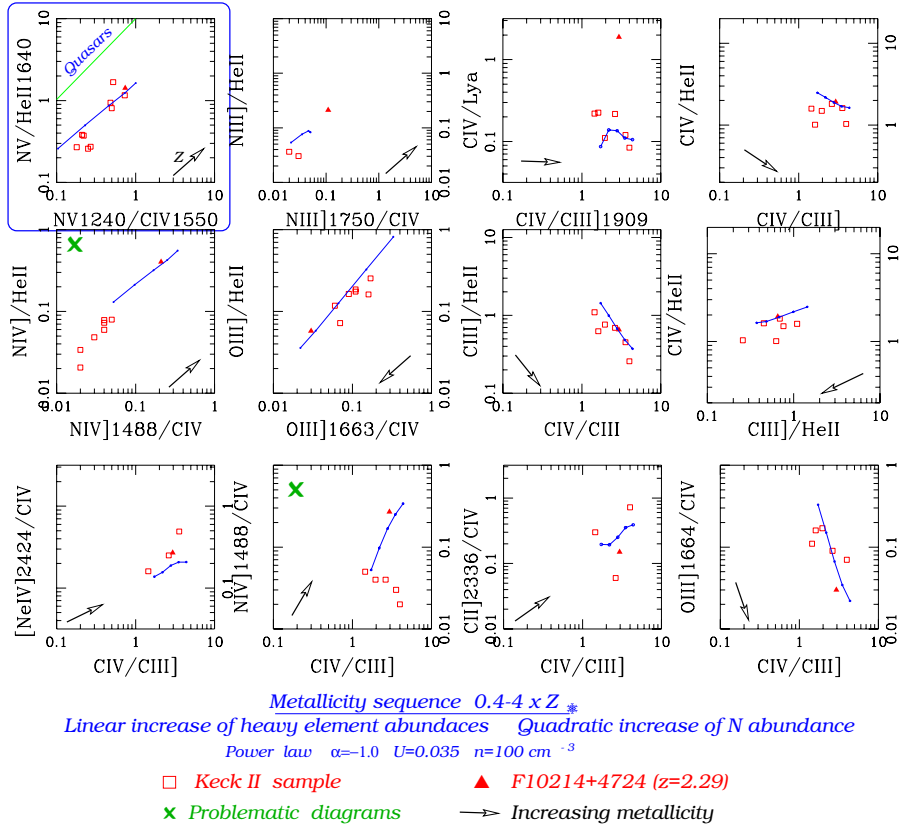}}
\caption{Diagnostic diagrams involving the strongest UV rest frame emission lines. 
The ``NV diagram'' is on the top left corner. The quasar correlation 
  is also shown. HzRG define a parallel correlation.
The  solid line is our metallicity sequence for the HzRG sample.
It shows  good agreement with the data in most diagnostic diagrams..}
\end{figure}

\section{Discussion and conclusions}

	Therefore, the  NV diagram suggests:

*  solar or supersolar metallicities
in the extended gas of many HzRG

* different levels of enrichment from object to object

* quadratic increase of N abundance, suggesting dominant secondary N production. 
This is consistent with studies  showing that secondary N production dominates 
at high metallicities \cite{hen00}.

	HF concluded from the NV diagram  that high redshift QSOs are associated
with vigorous star formation that enriches the gas in  short time
scales ($\leq$1 Gyr, at least for $z>$4 objects). 
Chemical evolution models  require a much faster evolution rate and a flatter
IMF compared to the solar neighbourhood case. 
The high abundances we derive require similar models.  This is 
the case of Giant Elliptical models (see \S6.2 in HF99) and this 
supports the idea
that HzRG (and quasars) are progenitors of giant ellipticals. Therefore, we conclude that also 
HzRG are undergoing  intense star formation
activity and we are witnessing the results of different evolutionary status
in different objects. 
The unification model for powerful radio galaxies and quasars \cite{bar89} supports our interpretation. 
	A more detailed discussion on the models and implications 
will be presented in Vernet
et al. (2001).

\end{article}

\end{document}